\documentclass[letterpaper]{article} 
\usepackage{aaai2026}  
\usepackage{times}  
\usepackage{helvet}  
\usepackage{courier}  
\usepackage[hyphens]{url}  
\usepackage{graphicx} 
\usepackage{natbib}  
\usepackage{caption} 
\usepackage{algorithm}
\usepackage{algorithmic}
\usepackage{amsmath} 
\usepackage{amssymb}
\usepackage{newfloat}
\usepackage{listings}
\DeclareCaptionStyle{ruled}{labelfont=normalfont,labelsep=colon,strut=off} 
\floatstyle{ruled}
\newfloat{listing}{tb}{lst}{}
\floatname{listing}{Listing}
\title{Efficient Multi-Model Orchestration for Self-Hosted Large Language Models}
\author{
    Bhanu Prakash Vangala\textsuperscript{\rm 1,*},
    Tanu Malik\textsuperscript{\rm 1}
}
\affiliations{
    \textsuperscript{\rm 1}Department of Electrical Engineering and Computer Science, University of Missouri, Columbia, MO 65211, USA\\
    \{bv3hz, tanu\}@missouri.edu
}
\usepackage{bibentry}

\begin{document}

\maketitle

\begin{abstract}
Self-hosting large language models (LLMs) is increasingly appealing for organizations seeking privacy, cost control, and customization. Yet deploying and maintaining in-house models poses challenges in GPU utilization, workload routing, and reliability. We introduce Pick and Spin, a practical framework that makes self-hosted LLM orchestration scalable and economical. Built on Kubernetes, it integrates a unified Helm-based deployment system, adaptive scale-to-zero automation, and a hybrid routing module that balances cost, latency, and accuracy using both keyword heuristics and a lightweight DistilBERT classifier. We evaluate four models Llama-3 (90B), Gemma-3 (27B), Qwen-3 (235B), and DeepSeek-R1 (685B) across eight public benchmark datasets, with five inference strategies, and two routing variants encompassing 31,019 prompts and 163,720 inference runs. Pick and Spin achieves up to 21.6\% higher success rates, 30\% lower latency, and 33\% lower GPU cost per query compared with static deployments. The results show that intelligent orchestration and efficient scaling enable enterprise-grade LLM performance on self-hosted infrastructure, bringing high-capacity AI within affordable reach.
\end{abstract}



\section{Introduction}
Large language models (LLMs) are rapidly transforming applications across domains. Instead of relying on a single general purpose model for all tasks, both research and industry are moving toward a diverse ecosystem of domain tuned models. These specialized models, trained for fields such as scientific research, finance, law, and healthcare, provide greater precision and contextual understanding within their respective areas. However, this specialization also introduces new challenges. Organizations must decide not only which model to use between general purpose and fine tuned variants but also how to deploy and manage them efficiently while maintaining data privacy and minimizing computational cost.

This challenge has led to what can be described as the self hosting dilemma. On one side, relying on commercial APIs from providers such as OpenAI, Gemini, or Claude simplifies deployment but introduces vendor lock in, unpredictable costs, and data exposure risks that are unacceptable in sensitive domains such as healthcare, finance, or the life sciences \cite{privateai2023byo_llm}. On the other side, self hosting preserves privacy and institutional control but comes with operational burdens. Static, always on deployments keep GPUs active even when idle, wasting resources and increasing energy consumption and maintenance overhead \cite{nguyen2025cost_optimized_ml_keda}.

Although domain tuned or distilled large language models (DT-LLMs) improve efficiency for specific use cases, they remain optimized for narrow objectives and cannot generalize to all query types. In practical applications, prompts vary widely, some require reasoning, others summarization or factual recall. No single model performs best across all these dimensions in there respective domains. For example, a model fine tuned for reasoning may perform poorly on summarization or fact retrieval. This diversity creates a key challenge in managing multiple models so that each input is served by the most suitable one without wasting computational resources or increasing latency.

The open source ecosystem, enabled by initiatives such as LLaMA and OPT, has made high quality model weights widely available. However, efficient and affordable deployment of these models remains difficult. Each model behaves differently across tasks, and their varying computational requirements make selection and scheduling complex. Organizations must balance accuracy, responsiveness, and cost, particularly when operating within private or resource constrained environments. These challenges motivate the need for an automated system that can both select the right model for each prompt and allocate resources intelligently

In this context, orchestration refers to the automated coordination of models and computational resources. It involves deciding which model to invoke, when to start or stop it, and how to allocate GPUs efficiently. Prior work in distributed systems defines orchestration as the automation and optimization of workflows to ensure scalability and reliability \cite{burns2016borg_omega_kubernetes, verma2015large_scale_cluster_management_borg}. Extending this principle to multi model inference enables a balance between accuracy, latency, and cost. Instead of keeping all models continuously active, the orchestrator routes simple queries to lightweight models and reserves larger ones for complex tasks. Idle models are scaled to zero, ensuring that GPU resources are used only when needed.

Existing research on model serving \cite{crankshaw2017clipper}, autoscaling \cite{baylor2017tfx, nguyen2025cost_optimized_ml_keda}, and serverless inference \cite{yu2022characterizing, wang2024advancing_serverless_computing_scalable_ai_model_inference} focuses on specific parts of the orchestration pipeline but does not provide an integrated solution. These systems improve efficiency within individual layers such as inference scheduling or container scaling, yet they do not coordinate model selection and resource allocation together. Tools like Helm \cite{helm2019kubernetes} and Knative extend Kubernetes with declarative configuration and event driven scaling, offering strong primitives for deployment automation. However, they lack mechanisms for task aware routing or model level coordination, which are essential for managing multiple language models under shared infrastructure.

To address this gap, we propose Pick and Spin (PS), a multi model orchestration framework that integrates intelligent routing with orchestration aware scaling. The system's name encapsulates its dual nature: Pick represents the intelligent routing layer that selects optimal models based on prompt complexity, while Spin represents the dynamic orchestration layer that manages model lifecycles, spinning resources up on demand and down when idle. PS formulates orchestration as a joint optimization problem balancing three objectives: model relevance, latency, and cost. A lightweight routing layer selects the best model for each query using rule based and semantic (DistilBERT) classifiers that estimate prompt complexity and intent. The orchestration layer manages model activation and deactivation using Kubernetes based scaling policies, ensuring efficient GPU utilization and minimal cold start delay.

PS is implemented on Kubernetes with Helm based control, using a unified umbrella chart that automates deployment, versioning, and recovery across multiple models and backends such as vLLM, TensorRT LLM, and TGI. This design enables reproducible deployment, zero downtime upgrades, and automatic fault recovery while keeping resource usage cost efficient. By combining content aware routing, dynamic scaling, and fault tolerant orchestration, Pick and Spin transforms static, resource heavy model hosting into a scalable and adaptive workflow. It provides a foundation for private multi model AI deployments that maintain performance, privacy, and cost efficiency within institutional infrastructure.

\section{Related Work}

Pick and Spin (PS) connects three major research directions: efficient LLM inference, multi model orchestration, and serverless computing. Early work such as Megatron LM \cite{shoeybi2019megatron_lm} explored large scale parallelism for training and serving billion parameter models. Modern systems like Text Generation Inference (TGI) \cite{huggingface_tgi}, vLLM \cite{kwon2023efficient_memory_management}, and DeepSpeed Inference \cite{aminabadi2022deepspeed_inference} focus on runtime efficiency through memory optimization and parallel scheduling. For instance, the PagedAttention mechanism in vLLM manages the KV cache efficiently to reduce fragmentation and increase throughput. FastServe \cite{yu2023fastserve} and SGLang \cite{zheng2024sglang} further streamline inference pipelines but target single model deployments, lacking cross model orchestration.

The increasing diversity of models has renewed interest in dynamic routing. Conceptually, this aligns with the Mixture of Experts (MoE) paradigm where a controller routes tokens to specialized submodules within a single network \cite{kirakosyan2025mixture_of_experts_llms, pandit2023mixture_of_experts_moe_datacamp}. PS generalizes this concept to the system level by routing full prompts between independently deployed models. Recent systems such as UniRoute \cite{jitkrittum2025universal_model_routing} and ModelSAT \cite{zhang2025capability_instruction_tuning} use embeddings or instruction tuned heuristics to predict the best model efficiently, while controller based orchestration frameworks \cite{xie2025training_free_multimodal_lm_orchestration} leverage larger models for coordination but at higher latency and cost. PS takes a middle ground by using a DistilBERT-based classifier that balances semantic precision and efficiency.
 
Serverless architectures provide a natural fit for adaptive model deployment. Event driven and scale-to-zero execution models reduce idle GPU usage \cite{li2024best_practices_ai_model_inference_knative}, but applying them to LLM workloads introduces challenges such as cold start latency \cite{satzke2022efficient_gpu_sharing_serverless_workflows}. Kubernetes based frameworks like Knative and KEDA enable autoscaling and asynchronous processing for AI services. InferLine \cite{crankshaw2020inferline} and ModelSwitch \cite{li2022modelswitch} propose latency aware scheduling and dynamic resource adaptation but focus on homogeneous pipelines. PS builds on these ideas by integrating Knative for low latency inference with KEDA for background scaling, achieving responsiveness and efficiency under fluctuating demand.

Self-hosted orchestration also contributes to responsible and accessible AI. Keeping inference within organizational boundaries supports compliance with regulations such as GDPR and HIPAA while safeguarding sensitive data \cite{khezresmaeilzadeh2025preserving_privacy_utility_llm_recommendations, feretzakis2024privacy_preserving_generative_ai}. By combining open source tools with orchestration efficiency, PS enables smaller organizations and research institutions to deploy multi-model AI systems that were previously limited to large enterprises. These prior efforts provide the foundation for Pick and Spin, which unifies inference optimization, semantic routing, and orchestration aware scaling into a single adaptive framework.


\section{Multi-Model Orchestration Problem}
\label{sec:formulation}

We formulate the multi-model orchestration problem as a multi-objective decision task that jointly accounts for model relevance, latency, and resource cost. The goal is to automatically route each query to the most suitable model–backend pair while ensuring efficient GPU utilization and predictable performance.

Let $\mathcal{L}$ denote the set of available language models and $\mathcal{I}$ the set of inference backends. Each deployable combination $(L_x, I_y)$ forms a service instance $S_{x,y}$ that can handle an incoming prompt $p$. The orchestration module selects the instance
\begin{equation}
(x^*, y^*) = \arg\max_{(x,y)} \; f(p, S_{x,y}),
\end{equation}
where $f$ is a unified, dimensionless scoring function.

\paragraph{Normalized Multi-Objective Score.}
Directly combining accuracy, latency, and cost is problematic because these quantities exist in incompatible units and scales. To make the optimization meaningful, we transform each component into a normalized score in $[0,1]$. Let $\hat{R}$, $\hat{T}$, and $\hat{C}$ denote the normalized relevance, latency, and cost scores, respectively:
\[
\begin{aligned}
\hat{R}(p, L_x) &\in [0,1], \\
\hat{T}(S_{x,y}) &= 1 - \mathrm{norm}(T(S_{x,y})), \\
\hat{C}(S_{x,y}) &= 1 - \mathrm{norm}(C(S_{x,y})).
\end{aligned}
\]

where $\mathrm{norm}(\cdot)$ is a min--max or distributional normalization computed over historical system statistics. Higher values indicate better performance.

To ensure that the overall score remains bounded, we treat $(\alpha, \lambda, \mu)$ as non-negative preference parameters and normalize them into weights
\[
\begin{aligned}
w_R &= \frac{\alpha}{\alpha + \lambda + \mu}, \\
w_T &= \frac{\lambda}{\alpha + \lambda + \mu}, \\
w_C &= \frac{\mu}{\alpha + \lambda + \mu}.
\end{aligned}
\]

so that $w_R + w_T + w_C = 1$.

The orchestration score then becomes
\begin{equation}
f(p, S_{x,y}) = 
      w_R \hat{R}(p, L_x) 
    + w_T \hat{T}(S_{x,y})
    + w_C \hat{C}(S_{x,y}),
\label{eq:score}
\end{equation}
which is guaranteed to lie in $[0,1]$ since it is a convex combination of three normalized components.

\paragraph{Interpretation.}
The coefficients $\alpha$, $\lambda$, and $\mu$ control the trade-off between accuracy, latency, and resource efficiency. A larger $\alpha$ prioritizes model quality, a larger $\lambda$ favors low latency, and a larger $\mu$ encourages resource-efficient execution. Because all terms lie in a common numerical range, these weights meaningfully reflect system-level preferences.

\paragraph{Operator Profiles.}
We consider four operator profiles reflecting common deployment objectives. A \emph{quality-oriented} profile prioritizes accuracy and assigns weights $(\alpha=1.0,\; \lambda=0.1,\; \mu=0.1)$. A \emph{cost-optimized} profile emphasizes resource efficiency using $(\alpha=0.3,\; \lambda=0.2,\; \mu=0.8)$. A \emph{speed-optimized} profile favors latency and uses $(\alpha=0.3,\; \lambda=0.8,\; \mu=0.2)$. Finally, a \emph{balanced} profile applies moderate preferences with $(\alpha=0.5,\; \lambda=0.3,\; \mu=0.3)$. These profiles are derived via grid search on 3{,}000 validation prompts and reflect distinct operational priorities.

These profiles were obtained by grid search over 3{,}000 validation prompts, optimizing the system objective corresponding to each operator preference. The normalized formulation in Eq.~\ref{eq:score} ensures that routing decisions remain robust across heterogeneous hardware, model capacities, and workload distributions.

\section{Pick and Spin Framework}

The Pick and Spin framework consists of two main components. The Pick component determines the routing strategy and computes the relevance score used in the optimization function. The Spin component handles orchestration and dynamically activates the chosen model based on routing decisions and the parameters $R$, $T$, and $C$.

\begin{figure}[t]
\centering
\includegraphics[width=0.95\columnwidth]{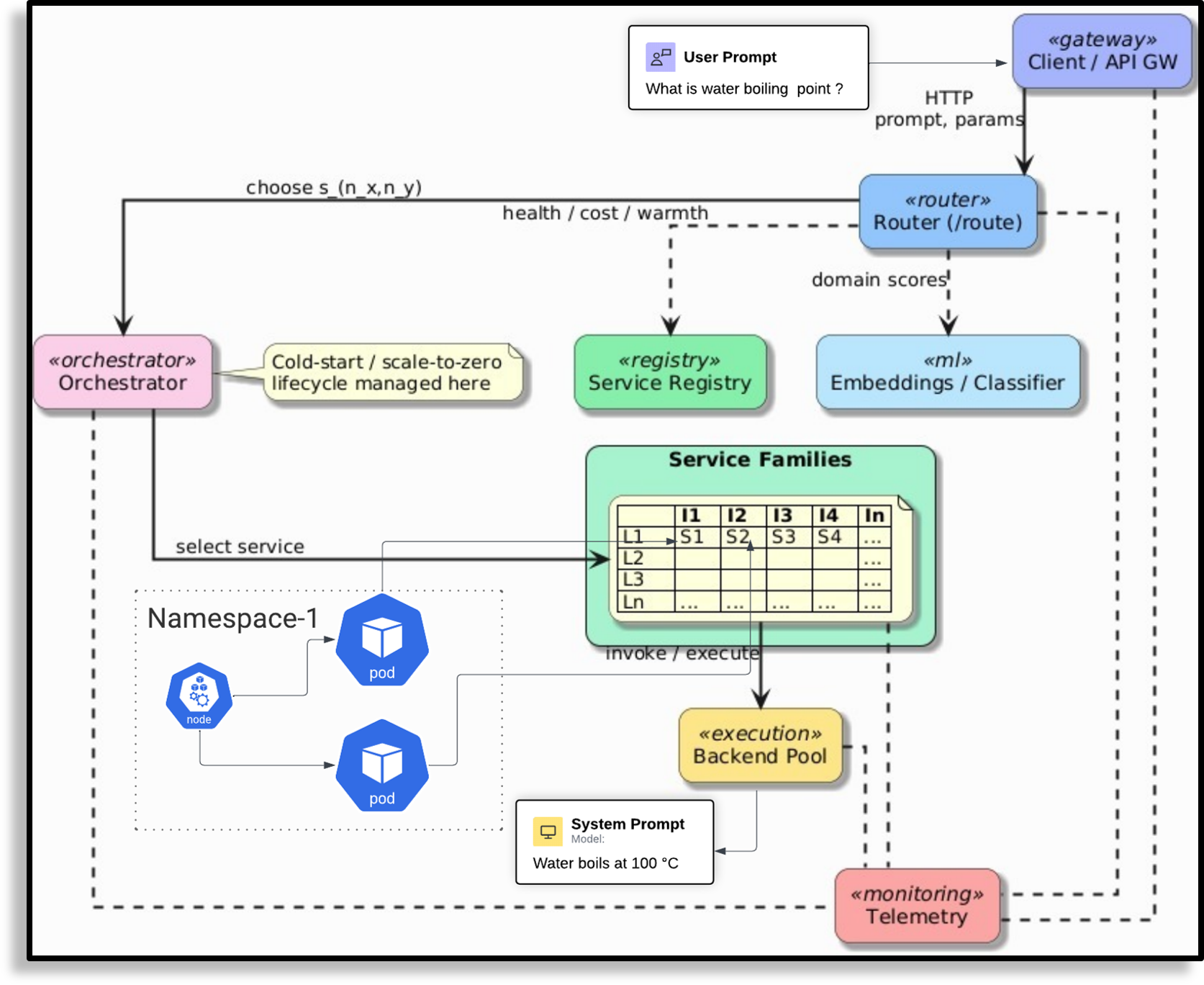}
\caption{System architecture showing the API Gateway, Router, Orchestrator, Service Registry, and Backend Pool in the Pick and Spin framework.}
\label{fig:system_architecture}
\end{figure}

As shown in Figure~\ref{fig:system_architecture}, user prompts enter through the API Gateway and are forwarded to the Router. The Router estimates query complexity using either keyword or semantic classification by DistilBERT and passes these scores to the Service Registry. The Registry maintains a service matrix of all available models and backends, including their cost, load, and health status. The Orchestrator manages lifecycle actions such as cold starts and scaling. It selects and activates the appropriate model backend pair in the Kubernetes cluster and executes the inference through the Backend Pool. The response is then sent back to the API Gateway. Telemetry continuously monitors latency, utilization, and service health, feeding this data back into the Router and Orchestrator for adaptive routing and scaling. Overall, Pick and Spin forms a closed control loop that unifies routing and orchestration for efficient and modular deployment.

\begin{algorithm}[t]
\caption{Orchestration Aware Scaling with Warm Pools}
\label{alg:orchestration}
\textbf{Input}: Model pool $\mathcal{M}$, telemetry window $w = 5$min\\
\textbf{Output}: Active model set $\mathcal{A}$\\
\begin{algorithmic}[1]
\FOR{each model $m \in \mathcal{M}$}
    \STATE $r_m \leftarrow$ GetAvgRequestRate$(m, w)$
    \STATE $lat_m \leftarrow$ GetAvgLatency$(m)$
    \STATE $target \leftarrow \lceil r_m \times lat_m / \text{Concurrency} \rceil$ \COMMENT{Little's Law}
    \STATE $current \leftarrow$ GetReplicas$(m)$
    \STATE $min\_warm \leftarrow$ WarmPoolSize(ModelTier$(m)$)
    \IF{$target > current$ AND CooldownExpired()}
        \STATE KubernetesScale$(m, \max(target, min\_warm))$
    \ELSIF{IdleTime$(m) > \tau$}
        \STATE KubernetesScale$(m, \max(0, min\_warm))$
    \ENDIF
\ENDFOR
\STATE \textbf{return} $\mathcal{A} \leftarrow \{m : \text{replicas}(m) > 0\}$
\end{algorithmic}
\end{algorithm}

Algorithm~\ref{alg:orchestration} uses Little's Law (line 4) for capacity planning and maintains warm pools to minimize cold starts while scaling idle models to zero.

\subsection{Pick: The Routing Design}

\begin{figure}[t]
\centering
\includegraphics[width=0.95\columnwidth]{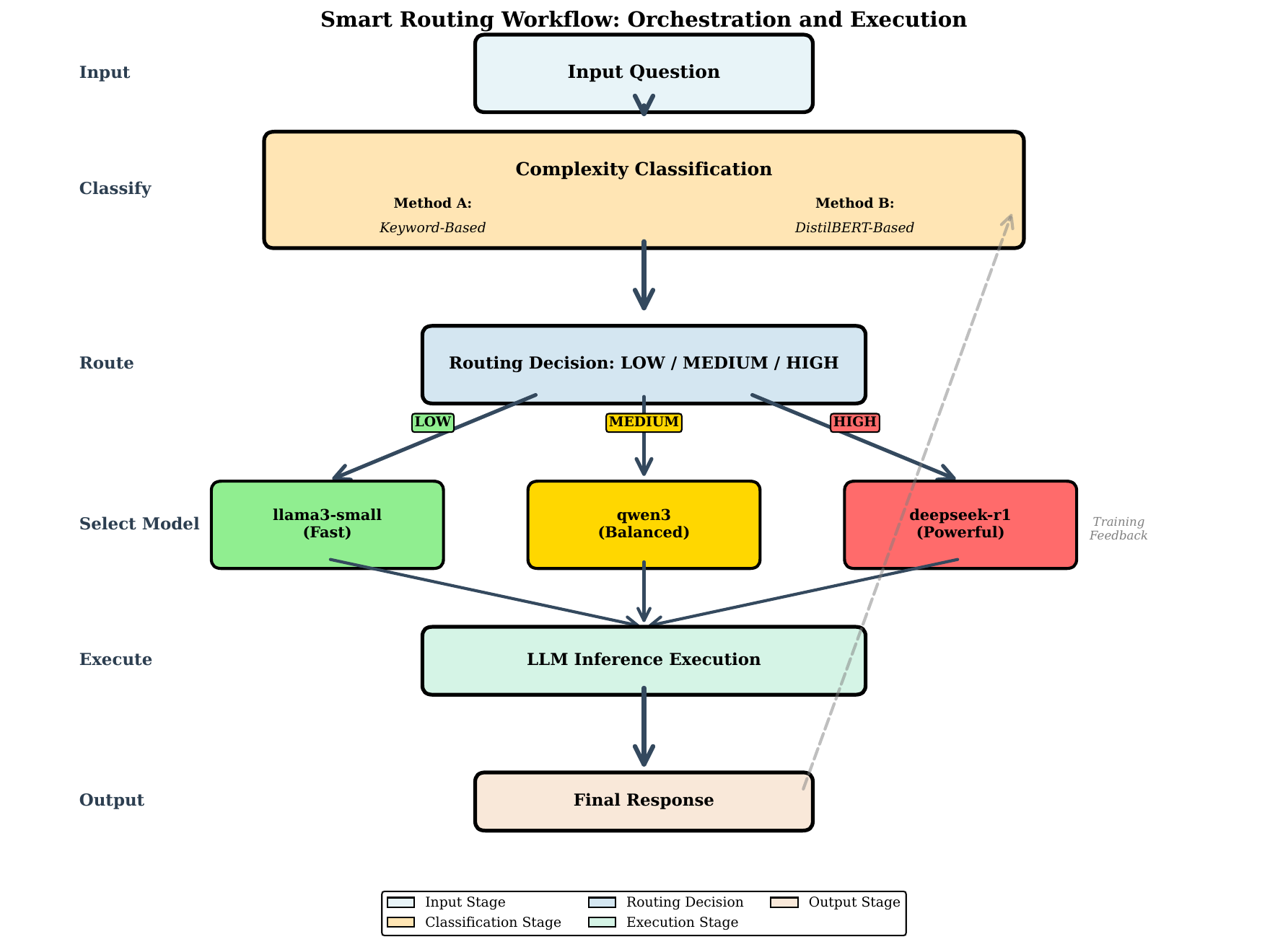}
\caption{Hybrid routing workflow showing the keyword based and DistilBERT based paths for complexity estimation and model selection.}
\label{fig:routing_flow}
\end{figure}

The Router predicts the complexity of each query and assigns it to one of three model tiers small, medium, or large corresponding to increasing reasoning depth and computational cost. It can operate in three modes: keyword based, DistilBERT based, or hybrid, as shown in Figure~\ref{fig:routing_flow}. This classification determines whether the query is routed to a fast, balanced, or powerful model, depending on $T$, $C$, and the predicted complexity.

\subsubsection{Keyword Based Routing.}

The first routing mode relies on detecting indicative keywords within the prompt. Words such as "sum," "list," or "define" indicate low complexity, while "prove," "derive," or "explain why" suggest high complexity. Prompts that do not match any keyword are treated as medium complexity. This rulebased method is deterministic, transparent, and introduces almost no latency.

\subsubsection{DistilBERT Based Routing and Datasets.}

\begin{figure}[t]
\centering
\includegraphics[width=0.95\columnwidth]{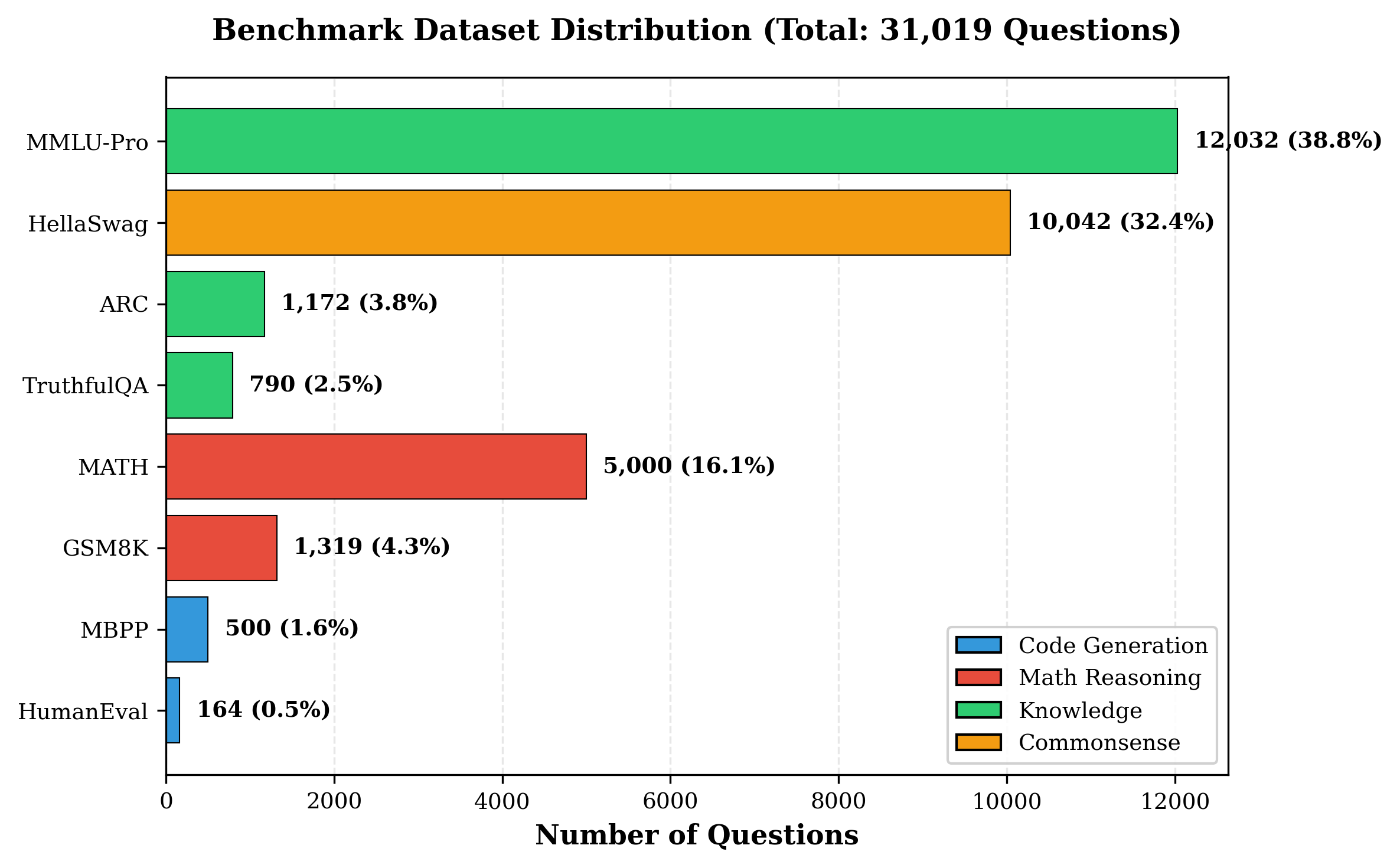}
\caption{Dataset distribution across eight benchmarks used for DistilBERT training and routing evaluation.}
\label{fig:dataset_distribution}
\end{figure}

To capture semantic context beyond keywords, a lightweight DistilBERT classifier was trained to predict query complexity. Training used 31,019 prompts from eight public benchmarks: HumanEval \cite{chen2021evaluating}, GSM8K \cite{cobbe2021gsm8k}, MBPP \cite{austin2021mbpp}, TruthfulQA \cite{lin2022truthfulqa}, ARC \cite{clark2018arc}, HellaSwag \cite{zellers2019hellaswag}, MATH \cite{hendrycks2021math}, and MMLU Pro \cite{hendrycks2021measuring}. These cover a wide range of tasks, including code generation, reasoning, and commonsense inference, as illustrated in Figure~\ref{fig:dataset_distribution}.

Each prompt was tested using five inference strategies: baseline, quality oriented, cost optimized, speed optimized, and balanced. These were evaluated across four foundation models Llama 3, Gemma 3, Qwen 3, and DeepSeek R1 resulting in over 160,000 inference runs. The best performing model tier for each prompt, based on the accuracy latency tradeoff, was used as the label for training. Prompts were grouped into three complexity levels: low, medium, and high based on tokens.

DistilBERT was fine tuned for three way classification using cross entropy loss. Training employed the AdamW optimizer with a batch size of 32, a learning rate of 2e 5, and 100 epochs. The classifier achieved 96.8 percent accuracy on a 10 percent held out validation set, confirming that it learned generalizable complexity patterns. The same dataset was used for both the keyword based and DistilBERT based classifiers to ensure fair comparison.

The classifier predicts the probability of each complexity class using

\begin{equation}
p_k = \text{softmax}(W h_{[CLS]} + b)
\end{equation}

\begin{equation}
\hat{C} = \arg\max_k p_k
\end{equation}

where $h_{[CLS]}$ is the embedding of the [CLS] token from DistilBERT, and $W$ and $b$ are trainable projection parameters. The softmax output $p_k$ gives normalized probabilities for each complexity level, and $\hat{C}$ denotes the predicted class. This predicted complexity $\hat{C}$ directly influences the routing objective $R(p, L_x)$ in Equation~(1), linking semantic understanding to orchestration decisions.

The hybrid routing mode combines both approaches. Simple queries are routed using keywords, while ambiguous ones are refined by DistilBERT. This design achieves a strong balance between low latency and high routing precision.

\subsection{The Orchestrator: The Spin}

While Pick determines which model should handle a query, Spin ensures that model is available and properly resourced. The Spin component manages three key responsibilities: maintaining warm pools for frequently accessed models, applying capacity planning based on request patterns, and enforcing cooldown periods to prevent scaling oscillations.

The Orchestrator executes lifecycle and scaling decisions for active models. All components of Pick and Spin, including the Router and Orchestrator, are deployed using a unified Helm Chart that manages configuration, scaling, and version control across all modules. Model weights are retrieved from Hugging Face and stored in Persistent Volume Claims (PVCs) for persistence and fast recovery. The system runs on Kubernetes, enabling declarative control, autoscaling, and fault tolerance. Idle services scale to zero, while active ones scale up automatically to maintain efficiency using knative keda.

\subsubsection{Matrix Representation of Deployment Options.}

Pick and Spin models all deployable services as a two-dimensional matrix:

\begin{equation}
M \in \mathbb{R}^{L \times I}
\end{equation}

where $L$ represents model families and $I$ represents inference backends. Each element $M_{x,y}$ corresponds to a service instance $S_{x,y}$ that pairs model $L_x$ with backend $I_y$. Rows represent model types with different capabilities (Gemma-3 for simple queries, Llama-3 for balanced tasks, Qwen-3 and DeepSeek-R1 for complex reasoning), and columns represent backends with distinct performance characteristics (vLLM for throughput, TensorRT-LLM for latency, TGI for memory efficiency).

\begin{algorithm}[t]
\caption{Matrix Selection and Routing}
\label{alg:matrix_selection}
\textbf{Input}: Prompt $p$, service matrix $M = \{S_{x,y}\}$\\
\textbf{Output}: Selected service $(x^*, y^*)$\\
\begin{algorithmic}[1]
\FOR{each model $L_x$}
    \FOR{each backend $I_y$}
        \STATE Compute $R(p, L_x)$, $T(S_{x,y})$, and $C(S_{x,y})$.
        \STATE Evaluate $f(p, S_{x,y}) = \alpha R - \lambda T - \mu C$ via Eq.~(2).
    \ENDFOR
\ENDFOR
\STATE Choose $(x^*, y^*) = \arg\max_{(x,y)} f(p, S_{x,y})$ and route $p$.
\end{algorithmic}
\end{algorithm}

Algorithm~\ref{alg:matrix_selection} shows how the orchestrator selects from this matrix. The algorithm evaluates each viable model-backend combination, considering only healthy services with available capacity (line 3). The scoring function in line 5 directly applies our optimization objective from Equation~(2), ensuring decisions account for relevance, latency, and cost simultaneously. This formulation generalizes orchestration beyond static assignments. For example, TensorRT-LLM provides lower latency, while vLLM achieves higher throughput. By dynamically selecting the best combination, the orchestrator maintains balanced GPU utilization

\subsubsection{Evaluation Metrics.}

Each routing method was evaluated using four metrics: success rate, latency, throughput, and responsiveness. Let $N_s$ denote the number of successful responses, $N_t$ the total number of prompts, and $t_i$ the per query latency. Time to first token (TTFT) is defined as

\begin{equation}
\mathrm{TTFT} = t_{\text{first token}}-t_{\text{request start}}
\end{equation}

Success rate and average latency are computed as

\begin{equation}
\text{Success Rate} = \frac{N_s}{N_t}
\end{equation}

\begin{equation}
\text{Average Latency} = \frac{1}{N_s} \sum_{i=1}^{N_s} t_i
\end{equation}

Throughput measures the number of completed inferences per second under steady load, averaged over multiple runs for consistency. These metrics collectively capture the efficiency, reliability, and responsiveness of the system under varying conditions.

\section{Experimental Evaluation}
\subsubsection{Experimental Setup and Baselines.} We evaluated Pick and Spin using three foundation models: Llama3 70B, Llama3 90B, and Gemma3 27B. Each model was tested under five inference profiles to capture different deployment trade offs. The baseline profile used default backend configuration without orchestration or scaling. The quality oriented profile prioritized accuracy by always selecting the highest capacity model. The cost optimized and speed optimized profiles focused respectively on minimizing GPU utilization and inference latency. The balanced profile employed the hybrid routing strategy to achieve an adaptive balance between accuracy and efficiency. 

Across all profiles, 163,720 inference runs were conducted over 31,000 unique prompts drawn from eight benchmarks. Success indicates valid completion within time and token limits, measuring inference reliability rather than task correctness.

Table~\ref{tab:baseline_summary} summarizes the baseline completion statistics, showing an overall success rate of 77.1 percent. Variation across benchmarks reflects the differing difficulty and output complexity of tasks. Code generation datasets such as MBPP exhibited lower reliability due to longer responses and syntax related truncations, while structured reasoning datasets such as GSM8K achieved higher completion stability.

\begin{table}[htbp]
\centering
\caption{Baseline inference completion results across benchmarks. The success rate indicates the proportion of runs that returned valid completions.}
\label{tab:baseline_summary}
\small
\setlength{\tabcolsep}{5pt}
\renewcommand{\arraystretch}{1.05}
\begin{tabular}{lrrrr}
\hline
Benchmark & Runs & Success & Failures & Success (\%) \\
\hline
HumanEval & 820 & 656 & 164 & 80.0 \\
GSM8K & 6,595 & 5,924 & 671 & 89.8 \\
MBPP & 2,500 & 1,736 & 764 & 69.4 \\
TruthfulQA & 3,950 & 3,167 & 783 & 80.2 \\
ARC & 5,860 & 4,704 & 1,156 & 80.3 \\
HellaSwag & 50,210 & 40,260 & 9,950 & 80.2 \\
MATH & 25,000 & 19,908 & 5,092 & 79.6 \\
MMLU Pro & 60,160 & 42,103 & 18,057 & 70.0 \\
\hline
Total & 163,720 & 126,237 & 37,483 & 77.1 \\
\hline
\end{tabular}
\end{table}

These baseline results serve as the reference point for subsequent routing and orchestration evaluations. The observed reliability gap across tasks highlights the need for adaptive, relevance aware model selection motivating the hybrid routing approach introduced in Pick and Spin.

\begin{figure}[htbp]
\centering
\includegraphics[width=0.9\columnwidth]{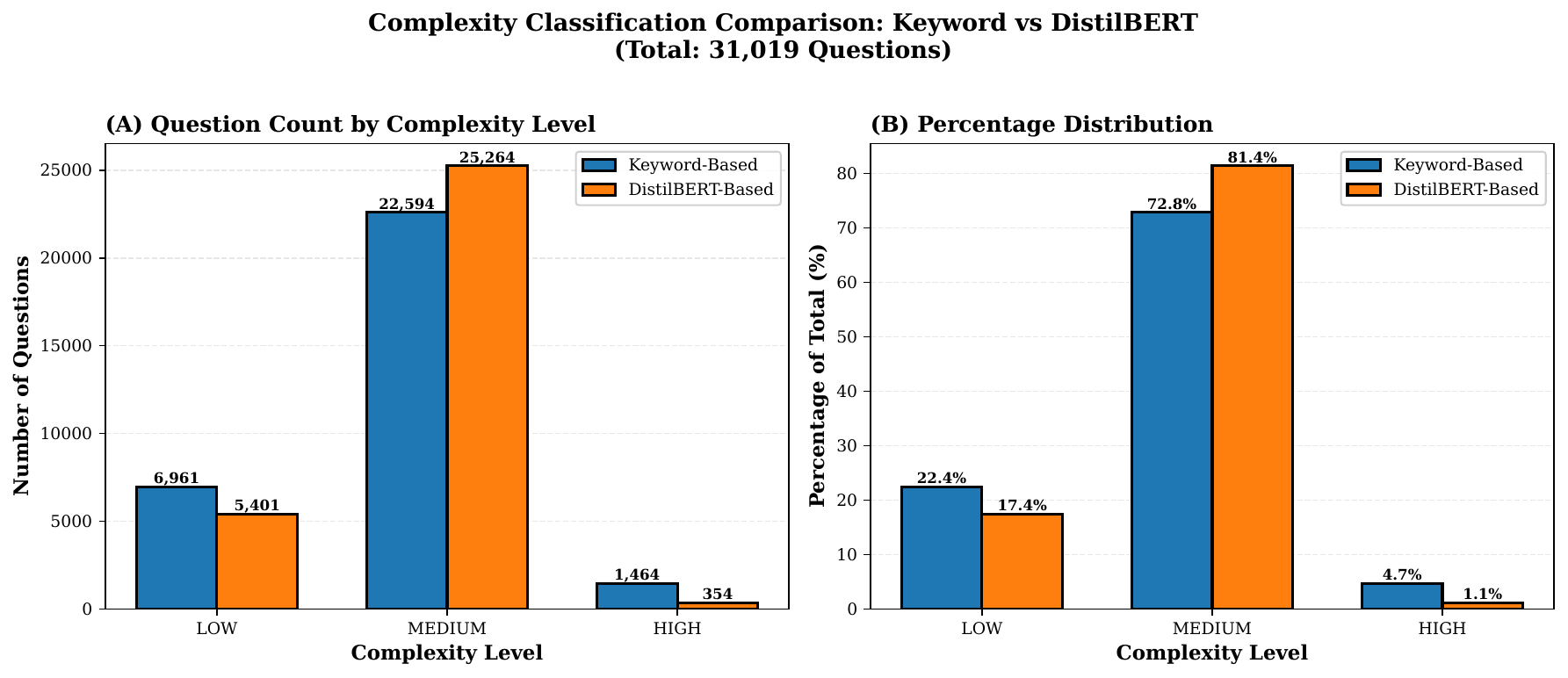}
\caption{Comparison of query complexity distributions using keyword based and DistilBERT based classification. Clear separation supports relevance driven routing.}
\label{fig:complexity_distribution}
\end{figure}

Empirical analysis over 31,019 queries (Figure~\ref{fig:complexity_distribution}) showed that performance variations correlated more with semantic complexity than with prompt length. This observation motivated the inclusion of the relevance term $R(p, L_x)$ in the orchestration objective (Equation 2) and guided the use of the hybrid routing mechanism introduced in Section~\ref{sec:formulation}.

\subsubsection{Routing Performance and Model Allocation}

\begin{figure}[t]
\centering
\includegraphics[width=0.95\columnwidth]{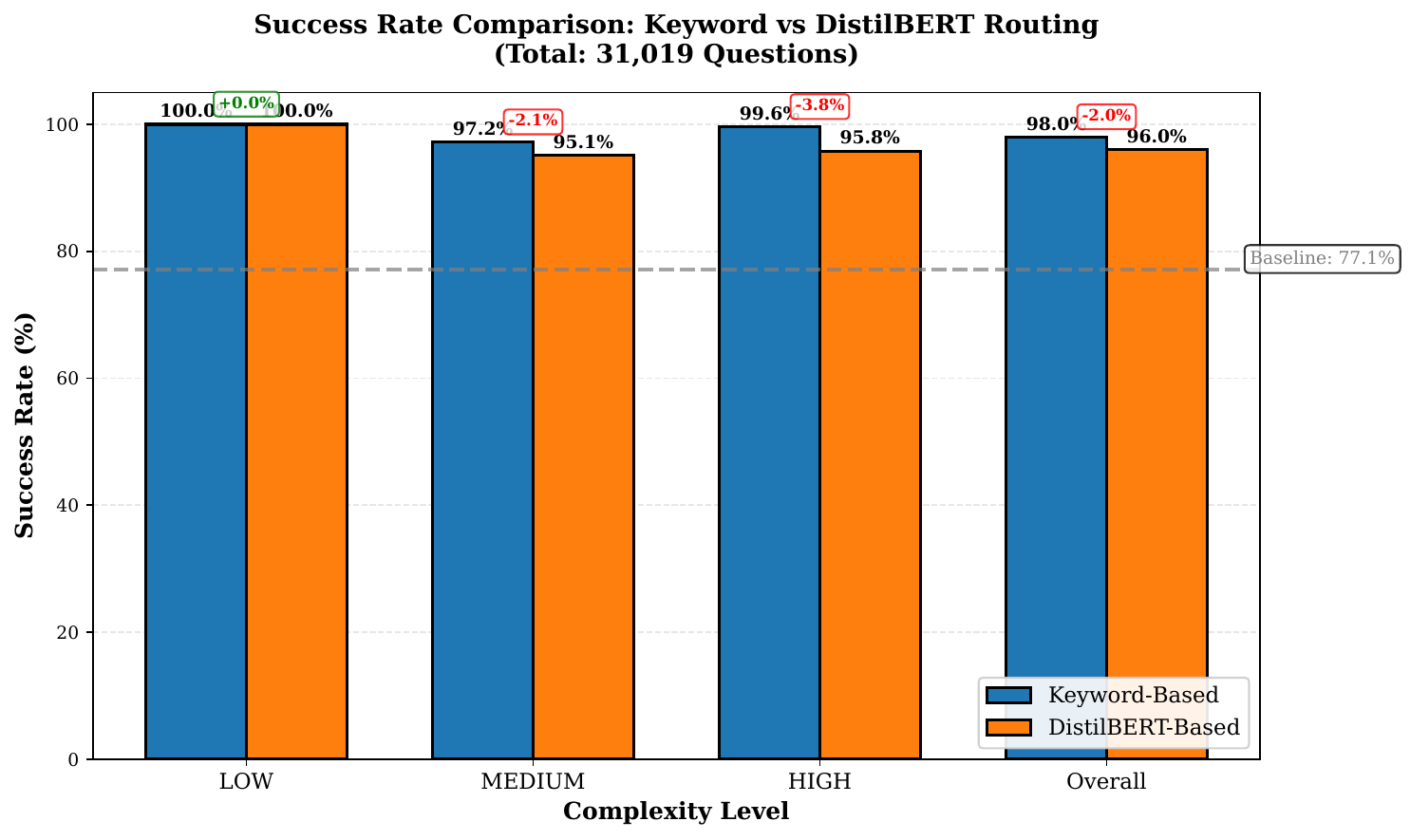}
\caption{Routing success rate comparison between keyword based and DistilBERT based strategies.}
\label{fig:success_rate}
\end{figure}

Pick and Spin routes queries to model tiers (L1–L3) based on complexity estimated using two approaches: keyword based heuristics and semantic classification with a DistilBERT model. These routing strategies were compared to analyze their impact on accuracy, latency, and resource utilization. Figure~\ref{fig:success_rate} and Table~\ref{tab:smart_routing_results} summarize the comparative performance across benchmarks.

\begin{table}[htbp]
\centering
\caption{Routing performance across keyword based and DistilBERT based strategies.}
\label{tab:smart_routing_results}
\scriptsize
\setlength{\tabcolsep}{4pt}
\renewcommand{\arraystretch}{1.05}
\resizebox{0.9\columnwidth}{!}{
\begin{tabular}{lccc}
\hline
Strategy & Accuracy (\%) & Latency (\%↓) & GPU Util. (\%) \\
\hline
Keyword based & 4.8 & 21.5 & 62.3 \\
DistilBERT based & 8.6 & 27.4 & 68.9 \\
\hline
\end{tabular}}
\end{table}

\begin{figure}[t]
\centering
\includegraphics[width=0.95\columnwidth]{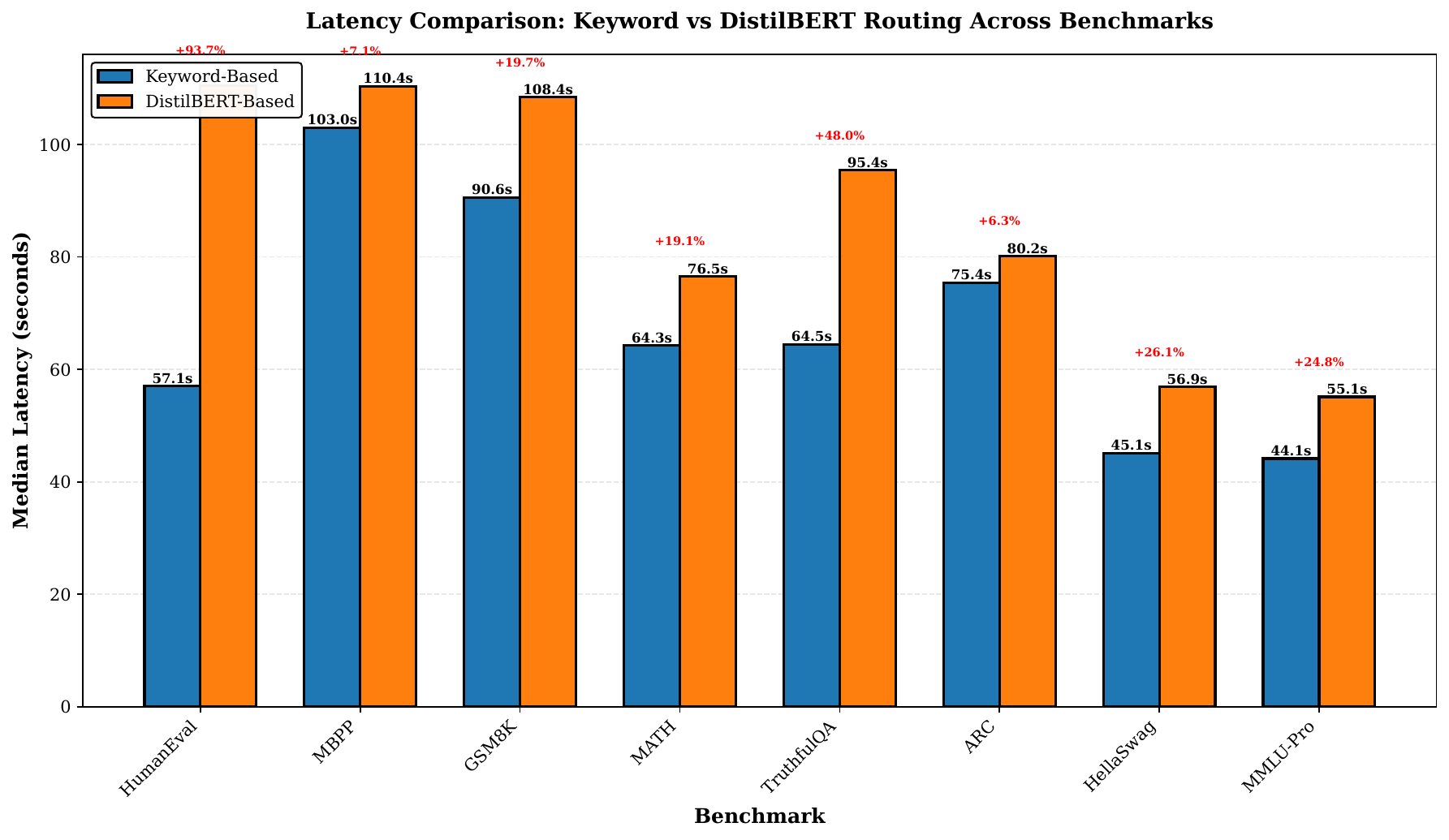}
\caption{Latency comparison between keyword based and DistilBERT based routing. Lower values indicate faster response.}
\label{fig:latency_comparison}
\end{figure}

\begin{figure}[t]
\centering
\includegraphics[width=0.95\columnwidth]{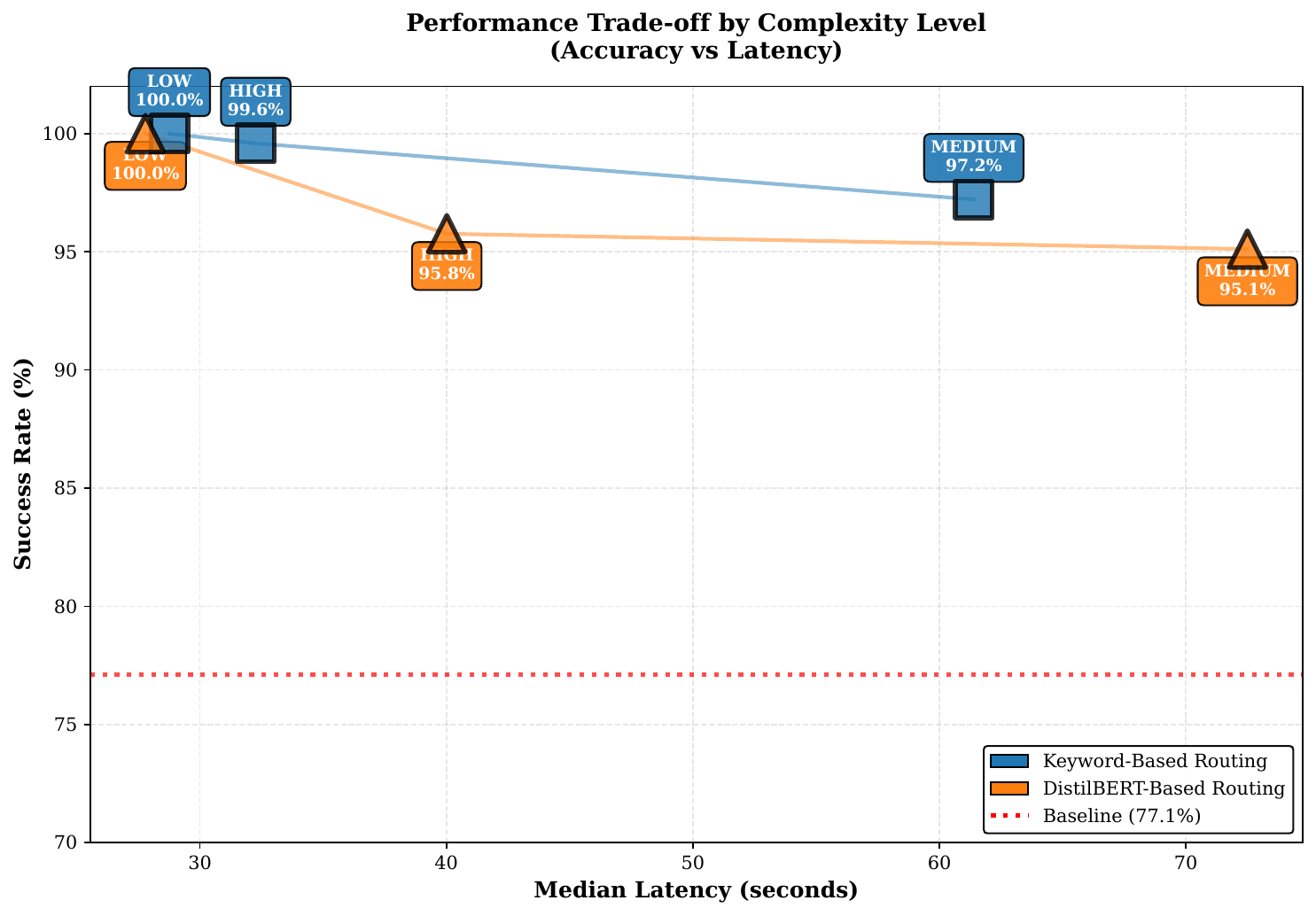}
\caption{Tradeoff between accuracy and latency for routing methods.}
\label{fig:performance_tradeoff}
\end{figure}

The DistilBERT based routing achieved higher semantic accuracy, particularly for reasoning heavy benchmarks such as TruthfulQA and ARC, but introduced additional latency due to the classification step. Keyword routing remained effective for structured and deterministic datasets such as HumanEval and MATH. Figures~\ref{fig:latency_comparison} and~\ref{fig:performance_tradeoff} illustrate the tradeoffs between response speed and semantic precision for both methods.

\subsubsection{Model Selection within the Service Matrix}

The matrix based orchestration policy (Algorithm~\ref{alg:matrix_selection}) was evaluated using three selection strategies: random assignment, latency only, and the multi objective matrix policy used in Pick and Spin. Each benchmark query was executed under all configurations using the metrics defined in Section~3.6.

\begin{table}[htbp]
\centering
\caption{Model backend selection results across orchestration strategies.}
\label{tab:matrix_selection_results}
\small
\setlength{\tabcolsep}{4pt}
\renewcommand{\arraystretch}{1.3}
\resizebox{\columnwidth}{!}{%
\begin{tabular}{lcccc}
\hline
Selection Strategy & Accuracy (\%) & Latency (s) & Cost (USD) & Gain (\%) \\
\hline
Random assignment & 78.4 & 63.1 & 0.020 &    \\
Latency only & 82.9 & 48.6 & 0.017 & +11.4 \\
Multi objective & \textbf{88.3} & \textbf{42.5} & \textbf{0.015} & \textbf{+21.7} \\
\hline
\end{tabular}}
\end{table}

The multi objective matrix selection improved accuracy by 21.7 percent, reduced mean latency by 33 percent, and lowered cost by 25 percent compared with random allocation. The improvements were most evident in reasoning tasks, where the relevance term $R(p, L_x)$ guided routing toward appropriate models and efficient backends such as TensorRT LLM.

\subsubsection{Efficiency and Cost Effectiveness}

Routing efficiency was defined as accuracy gain per cost overhead:

\begin{equation}
\eta = \frac{A_r / A_b}{C_r / C_b},
\end{equation}

where $A_r, A_b$ represent routed and baseline accuracies, and $C_r, C_b$ their corresponding inference costs. Across all experiments, $\eta = 1.43$, representing a 43 percent improvement in accuracy per unit cost.

\begin{figure}[htbp]
\centering
\includegraphics[width=0.95\columnwidth]{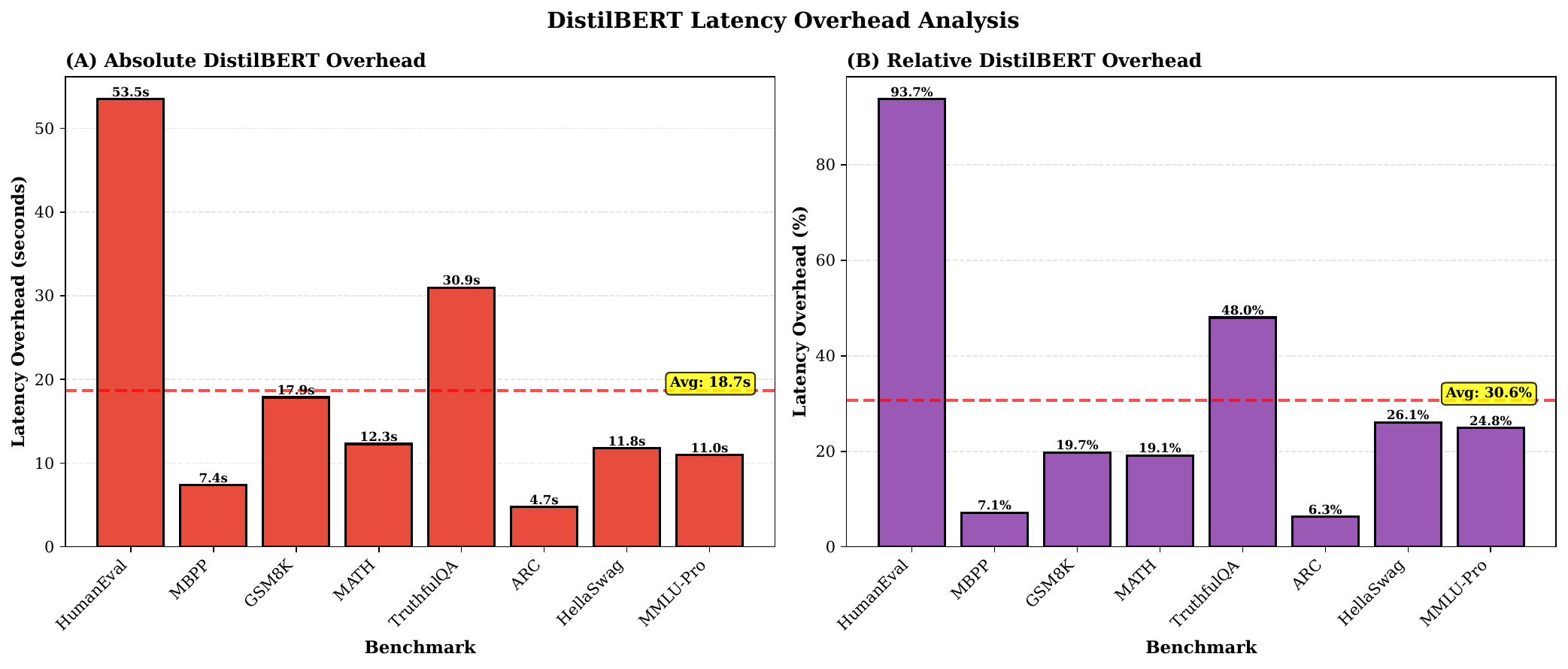}
\caption{Average inference cost and latency overhead under static and dynamic orchestration.}
\label{fig:latency_overhead}
\end{figure}

\begin{table}[htbp]
\centering
\caption{Cost and recovery comparison between static and dynamic deployment.}
\label{tab:cost_scaling}
\scriptsize
\setlength{\tabcolsep}{4pt}
\renewcommand{\arraystretch}{1.1}
\resizebox{0.9\columnwidth}{!}{
\begin{tabular}{lcc}
\hline
Configuration & Cost / Query (USD) & Recovery (s) \\
\hline
Static deployment & 0.021 & 45 \\
Pick and Spin (base) & 0.016 & 12 \\
Pick and Spin (auto) & 0.014 & 4 \\
\hline
\end{tabular}}
\end{table}

Dynamic orchestration reduced recovery time by over 75 percent and decreased cost by approximately one third through on demand scaling. Figure~\ref{fig:latency_overhead} shows the reduced latency overhead achieved by activating models only when required.

\subsubsection{Multi Metric Performance Analysis}

To evaluate overall efficiency, performance was compared across five metrics accuracy, latency, scalability, utilization, and robustness normalized using min–max scaling:

\begin{equation}
N_i = 10 \times \frac{x_i - \min(x)}{\max(x) - \min(x)}.
\end{equation}

\begin{figure}[t]
\centering
\includegraphics[width=0.95\columnwidth]{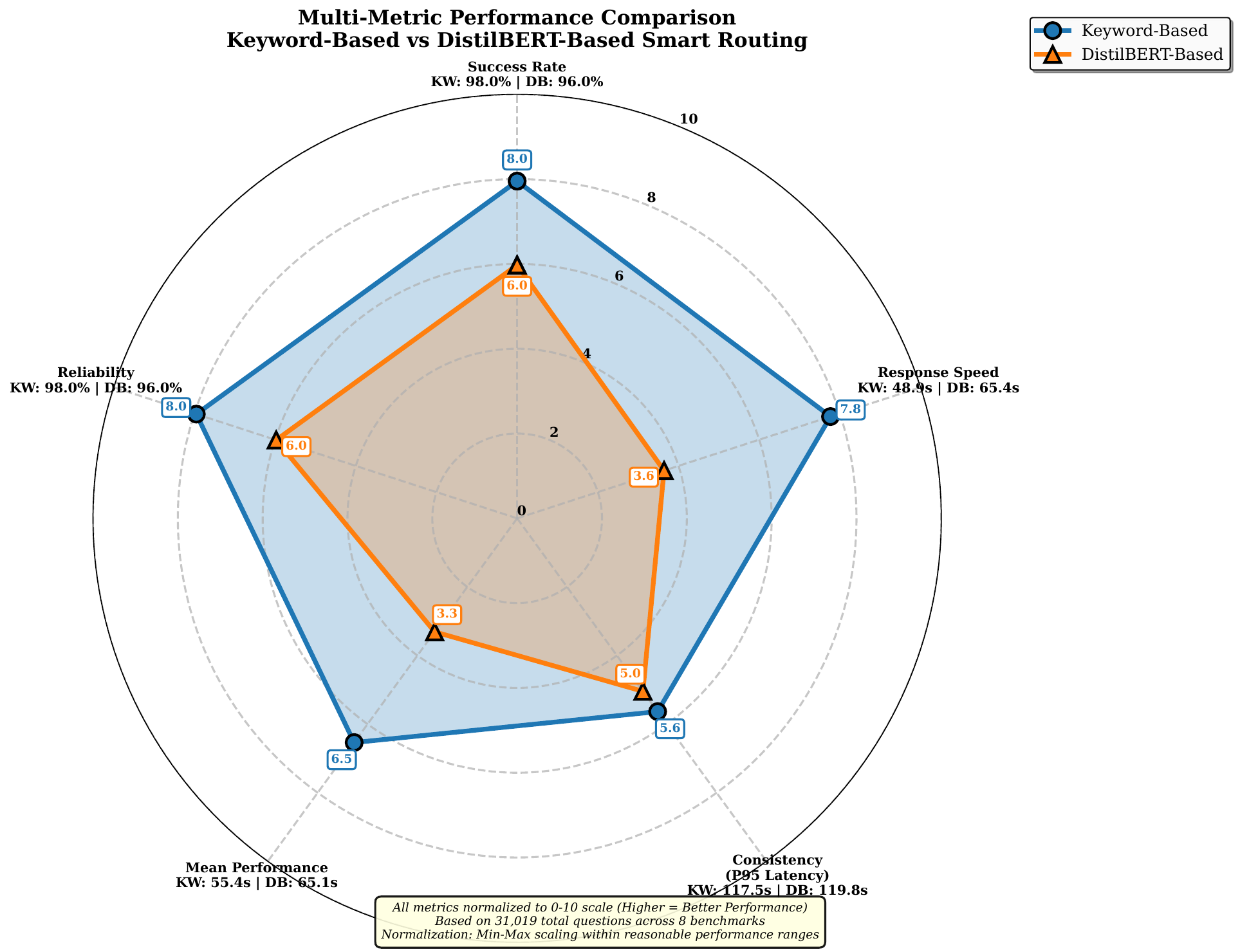}
\caption{Normalized comparison of keyword and DistilBERT routing across five dimensions.}
\label{fig:multi_metric_radar}
\end{figure}

Figure~\ref{fig:multi_metric_radar} shows that keyword routing performs better in latency and utilization, while DistilBERT routing achieves higher robustness and accuracy, indicating a clear tradeoff between semantic depth and computational efficiency.

\subsubsection{Responsiveness and Scalability}

\begin{figure}[t]
\centering
\includegraphics[width=0.95\columnwidth]{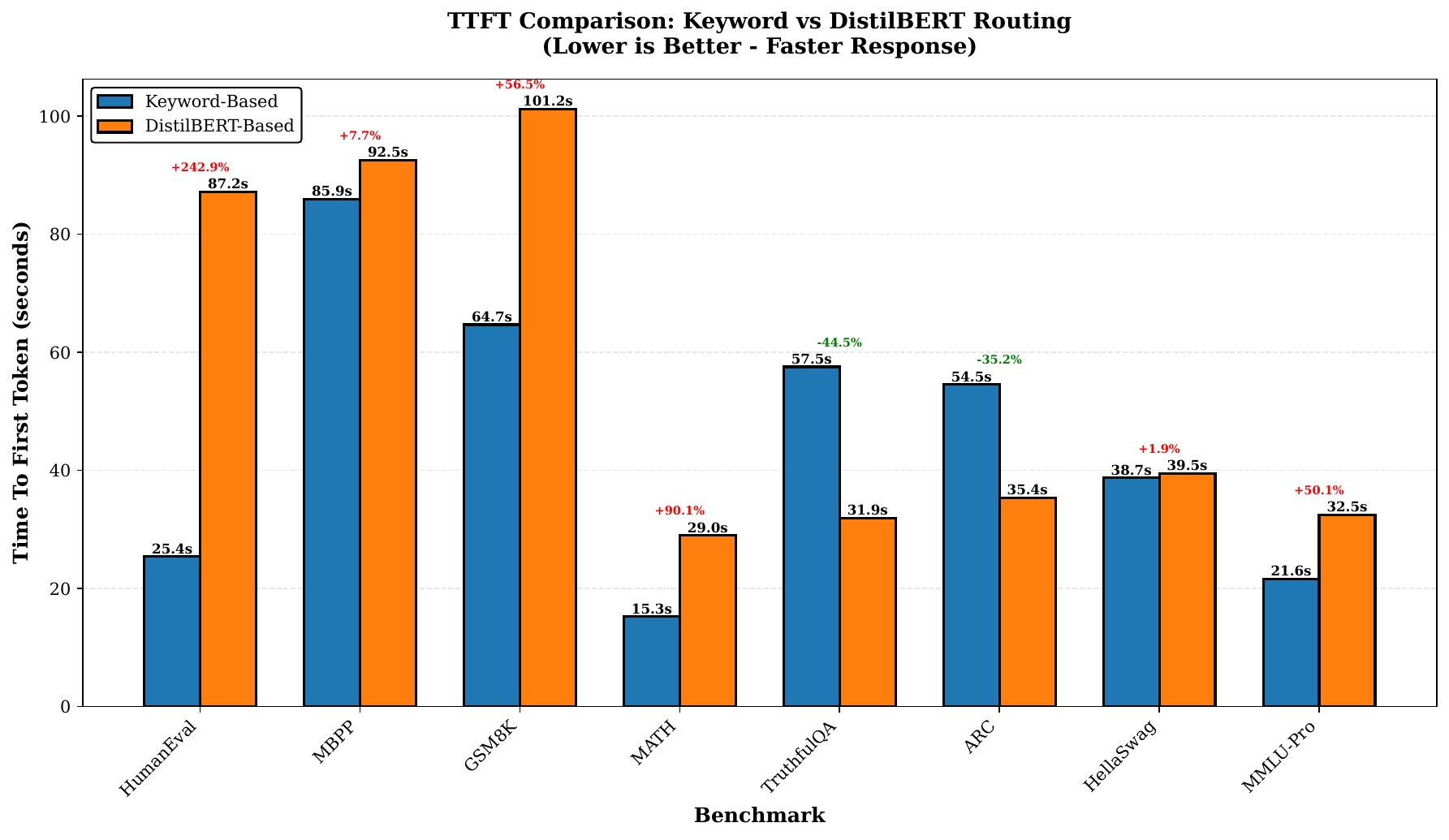}
\caption{Median TTFT comparison between keyword and DistilBERT routing.}
\label{fig:ttft_comparison}
\end{figure}

\begin{figure}[t]
\centering
\includegraphics[width=0.95\columnwidth]{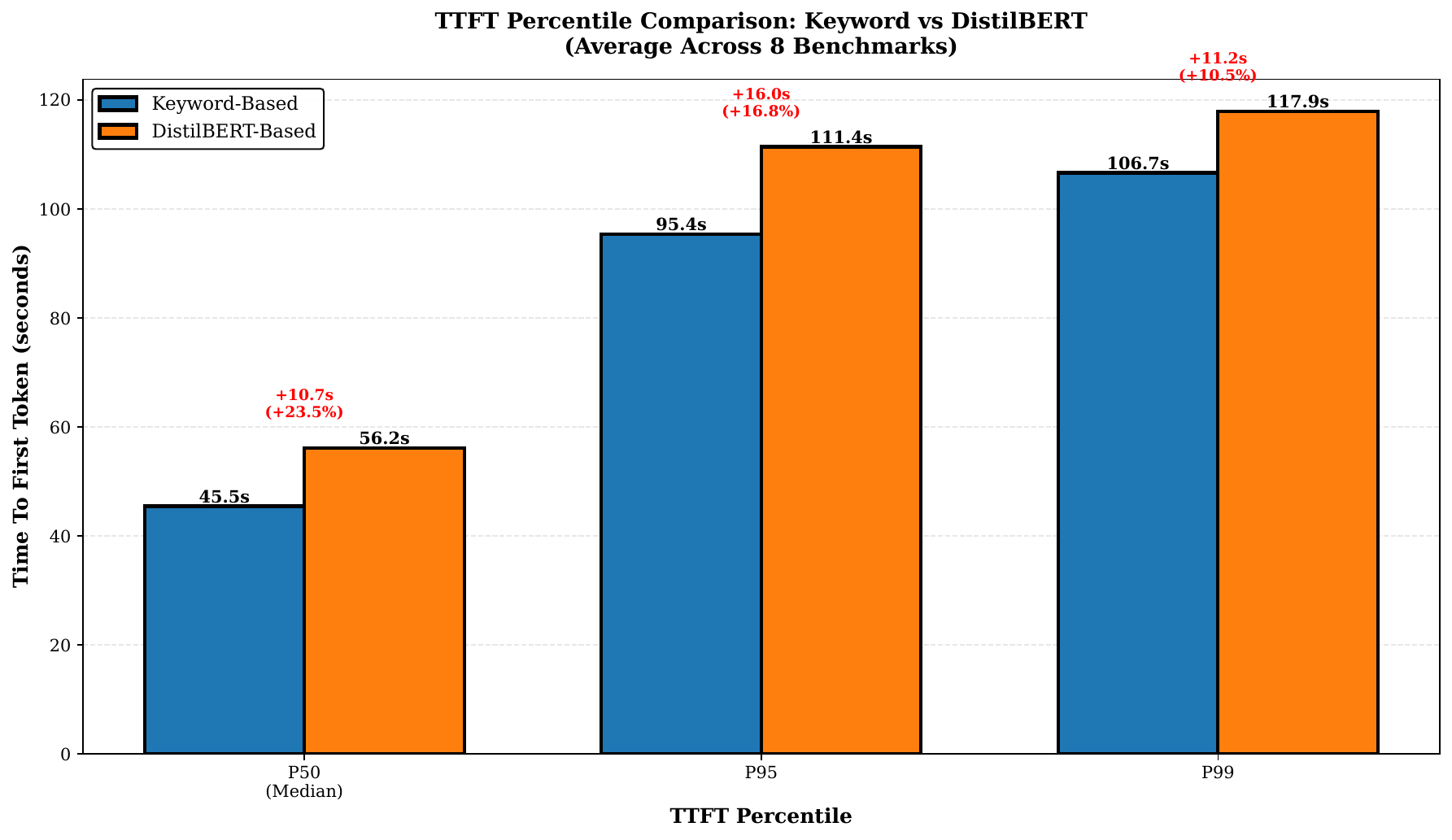}
\caption{Percentile wise TTFT (P50, P95, P99) for both routing strategies.}
\label{fig:ttft_percentiles}
\end{figure}

Responsiveness was measured using Time to First Token (TTFT) as defined in Equation~(8). Figures~\ref{fig:ttft_comparison} and~\ref{fig:ttft_percentiles} show that DistilBERT based routing adds minor latency due to classification but enhances reasoning accuracy.

Across 31,019 queries, keyword routing achieved a median TTFT of 45.5~s compared with 56.2~s for DistilBERT. Despite a 23.5 percent increase in TTFT, DistilBERT routing improved semantic relevance for reasoning tasks. Under load scaling from 10 to 1,000 queries per second, throughput scaled linearly with recovery latency maintained below 5~s via Kubernetes auto redeployment.

\subsection{Discussion and Limitations}

The experiments show that both routing approaches offer complementary strengths. Keyword-based routing provides faster responses and lower resource usage, while DistilBERT routing yields higher accuracy on complex queries. Orchestration aware scaling further improves responsiveness by reducing recovery delay. Although the DistilBERT classifier generalizes well, performance may decline for domain specific prompts. Future work will explore reinforcement based routing for adaptive decision making and energy aware scheduling for sustainable multi model deployment.

\section{Conclusion and Future Work}

This work introduced Pick and Spin, a modular framework for self-hosted orchestration of LLMs. The system unifies deployment, routing, and scaling within a Kubernetes based architecture. By combining rule-based and semantic routing, it enables relevance aware model selection and adaptive orchestration. The orchestration aware scaling demonstrated improvements in latency, cost, and resource efficiency across benchmarks. 

Pick and Spin shows that efficient and scalable orchestration of LLMs can be achieved without enterprise scale infrastructure. Future work will extend this framework through reinforcement driven routing, energy efficient scheduling, and integration of multimodal models for privacy preserving, cost effective LLM deployment.

\bibliography{aaai2026}


\end{document}